# On the Exact Maxwell evolution equation of resonator dynamics

**Tong Wu,[1] Rachid Zarouf,[2,3] Philippe Lalanne[1,*]**

[1]Laboratoire Photonique, Numérique et Nanosciences (LP2N), IOGS- Université de Bordeaux-CNRS, 33400 Talence cedex, France
[2]Aix-Marseille Université, Laboratoire ADEF, Campus Universitaire de Saint-Jérôme, 52 Avenue Escadrille Normandie Niémen, 13013 Marseille, France
[3]CPT, Aix-Marseille Université, Université de Toulon, Marseille, France
*philippe.lalanne@institutoptique.fr*

**Abstract:** In a recent publication [Opt. Express **32**, 20904 (2024)], the accuracy of the main evolution equation that governs resonator dynamics in the coupled-mode theory (CMT) was questioned. The study concluded that the driving force is proportional to the temporal derivative of the excitation field rather than the excitation field itself. This conclusion was reached with a derivation of an "exact" Maxwell evolution (EME) equation obtained directly from Maxwell's equations, which was further supported by extensive numerical tests. Hereafter, we argue that the original derivation lacks mathematical rigor. We present a direct and rigorous derivation that establishes a solid mathematical foundation for the EME equation. This new approach clarifies the origin of the temporal derivative in the excitation term of CMT and elucidates the approximations present in the classical CMT evolution equation through a straightforward argument.

## 1. Introduction

Many concepts in physics are based on normal modes of conservative systems, such as molecular orbitals and excitation energies. In these self-adjoint problems, the response of the system can be represented as a sum over the complete set of normal modes.

However, when energy dissipation occurs through processes like absorption or leakage in open systems, the system ceases to be conservative. The corresponding operator becomes non-self-adjoint and its spectrum becomes continuous. The spectrum also includes an infinite set of quasinormal modes (QNMs) characterized by complex eigenfrequencies (or eigenenergies), which produce characteristic damped oscillations in the system's temporal response. A significant challenge across various disciplines [1-4] focuses on using QNM expansions to represent this response, akin to the methods applied to closed systems with normal modes.

In electromagnetism, this challenge was effectively addressed long ago with the temporal coupled-mode theory (CMT) [5-6]. CMT offers a computationally effective, systematic, and intuitive framework for characterizing interactions between different modes within resonators and predicting their temporal dynamics. As a result, it has become an essential tool for designing and optimizing optical resonators, facilitating the development of innovative photonic technologies in both linear [3,4,7] and nonlinear optics [8].

Typically, CMT assumes the presence of $m$ modes and $n$ ports. In this context, a mode corresponds to a QNM, while a port represents a propagating channel where the QNMs may decay. It is perhaps worth noting that our understanding of ports may not fully reflect reality, as resonators are open systems that dissipate energy across a continuum of "ports". A typical scenario involves a resonator on a layered substrate, illuminated by a plane wave (one of the system's many ports). This wave scatters into a continuum of ports composed of all the plane waves in the substrate or superstate, as well as any potential guided modes of the layered substrate [9].

The CMT equations consist of two primary equations [5-8]. The first, known as the evolution equation, describes how the amplitudes of the modes evolve in response to incoming waves from the ports. This equation is crucial for understanding the dynamics of the resonator. The second equation coherently combines resonant-assisted and background pathways to predict the fields that couple out into the ports, enabling the calculation of reflection and transmission coefficients for the various ports.

Despite its extensive application and many successes, CMT remains a phenomenological theory for Fano-Feshbach resonances [10]; all coefficients describing the coupling between the resonator modes and the ports are fitted from experimental or numerical data. There are exceptions for certain geometries. For instance, for microring filters, the coupling issue may accurately reduce to calculating the interaction between guided modes in bent waveguides with a Hermitian theoretical framework [11].

In this work, we concentrate on the first equation and its theoretical derivation. In a recent study [10], by trying to establish a rigorous foundation for the CMT using electromagnetic QNM theory, an alternative evolution equation, distinct from the one proposed in the CMT, was derived directly from Maxwell equations. This new equation, termed "exact" Maxwell Evolution (EME) equation, shares some similarities with the classical CMT evolution equation but also shows notable differences. A key distinction is in the nature of the driving term: in the EME equation, the excitation is proportional to the temporal derivative of the incident field, whereas in the CMT evolution equation, it is proportional to the incident field itself. This unexpected result was validated through a numerical test, which remains unchallenged in this study. Additionally, it was supported by a mathematical demonstration.

Here, we highlight that the demonstration lacks from mathematical rigor and propose a new demonstration that is both direct and rigorous. Importantly, the new demonstration also helps clarifying the key difference between a driving force proportional to the incident field and one proportional to its temporal derivative.

The manuscript is organized as follows. Section 2 helps the reader in contextualizing the topic by reintroducing the origin of the double integral necessary for deriving the EME equation. Section 3 highlights several aspects of the demonstration in Section 2 of [10] which lack mathematical rigor. Section 4 provides a direct and explicit demonstration of the EME equation, which is further supported in Appendix A by a more general demonstration with minimum assumptions on the incident field. Section 5 clarifies the approximation carried out by assuming a driving force directly proportional to the incident field. Section 6 concludes the work.

## 2. Background

We start by considering recent results obtained for QNM-expansions in the spectral domain for harmonic fields. Electromagnetic QNMs, labeled by the integer $m$, are source-free solutions of Maxwell equations, $\nabla \times \tilde{\mathbf{E}}_m = -i\tilde{\omega}_m \mu_0 \tilde{\mathbf{H}}_m$, $\nabla \times \tilde{\mathbf{H}}_m = i\tilde{\omega}_m \boldsymbol{\varepsilon}(\tilde{\omega}) \tilde{\mathbf{E}}_m$, which satisfy the outgoing-wave condition for $|\mathbf{r}| \to \infty$ [3,4]. Here, $\tilde{\mathbf{E}}_m$ and $\tilde{\mathbf{H}}_m$ represent the QNM electric and magnetic fields, respectively, $\boldsymbol{\varepsilon}$ denotes the possibly dispersive permittivity tensors. The QNM fields decay exponentially over time, so that $\text{Im}(\tilde{\omega}_m) < 0$ in the $\exp(-i\omega t)$ notation. From this point forward, we assume the QNMs are normalized, which is a crucial step in any QNM theory. There are several methods available for normalizing electromagnetic QNMs, all of which are thoroughly discussed in [12]. In this work, we utilize the so-called PML normalization approach, where PML stands for Perfectly Matched Layer.

In the scattered-field formulation, the total field at frequency $\omega$ is decomposed as a sum of the background field $[\mathbf{E}_b(\mathbf{r},\omega), \mathbf{H}_b(\mathbf{r},\omega)]\exp(-i\omega t)$ and the field $[\mathbf{E}_s(\mathbf{r},\omega), \mathbf{H}_s(\mathbf{r},\omega)]\exp(-i\omega t)$ scattered by the nanoresonator. The scattered field can be expanded in an infinite set of normalized QNMs and PML modes [12,13]

$$[\mathbf{E}_s(\mathbf{r},\omega), \mathbf{H}_s(\mathbf{r},\omega)] = \sum_m \alpha_m(\omega)[\tilde{\mathbf{E}}_m(\mathbf{r}), \tilde{\mathbf{H}}_m(\mathbf{r})], \qquad (1)$$

where $\alpha_m(\omega)$ denotes the excitation coefficient of the $m^{th}$ modes. Note that the normalized modes satisfy $\int \left[\tilde{\mathbf{E}}_m \cdot \left(\frac{\partial \omega \varepsilon}{\partial \omega}\right) \tilde{\mathbf{E}}_m - \tilde{\mathbf{H}}_m \cdot \left(\frac{\partial \omega \mu_0}{\partial \omega}\right) \tilde{\mathbf{H}}_m\right] d^3r = 1$ [12]. The scattered and normalized mode fields thus have different units.

The QNM framework based on a complex-mapping regularization provides a unique expression for the excitation coefficient in Eq. (1) [13]

$$\alpha_m(\omega) = \frac{\omega}{\widetilde{\omega}_m - \omega} \int_{V_r} \Delta\varepsilon(r) \mathbf{E}_b(r,\omega) \cdot \widetilde{\mathbf{E}}_m d^3 r, \tag{2}$$

for the case of nondispersive materials under consideration here. An equivalent EME equation for dispersive systems with Drude-Lorentz permittivities is derived in [13]. In this case, the driving force consists of two components: one proportional to the temporal derivative of the excitation field, and the other to the excitation field itself. Numerical tests conducted on a silver bowtie antenna, as shown in Fig. 2b of [13], strongly validate the robustness of the analysis. However, it would be valuable to expand the study further, for example, by conducting additional numerical tests for different geometries. It would be also valuable to generalize the theory to materials with arbitrary dispersion.

Equation (2) corresponds to Eq. (3) in [10]. $\alpha_m(\omega)$ essentially represents an overlap integral between the QNM and the background field. $\Delta\boldsymbol{\varepsilon}$ specifies the permittivity variation used for the scattered-field formulation, given by $\Delta\boldsymbol{\varepsilon}(r) = \boldsymbol{\varepsilon}_R(r) - \boldsymbol{\varepsilon}_b(r)$, where $\boldsymbol{\varepsilon}_R(r)$ and $\boldsymbol{\varepsilon}_b(r)$ denote the permittivities of the resonator and the background, respectively. In general, $\Delta\boldsymbol{\varepsilon}(r)$ is non-zero within a compact volume $V_R(r)$ that defines the resonator in the scattered-field formulation.

We now consider that the background field is an optical pulse, $\mathbf{E}_b(r,t)$, i.e. a wave packet that can be Fourier transformed $\mathbf{E}_b(r,\omega) = (2\pi)^{-1} \int_{-\infty}^{\infty} \mathbf{E}_b(r,t) \exp(i\omega t)\, dt$. Driven by the incident pulse, the resonator scatters a time-dependent electric field, $\mathbf{E}_s(r,t)$. Every infinitesimal frequency component of the background field $\mathbf{E}_b(r,\omega)d\omega$ gives rise to an infinitesimal harmonic scattered field $d\mathbf{E}_s(r,\omega) = \sum_m \alpha_m(\omega) \widetilde{\mathbf{E}}_m(r)\, d\omega$, and the scattered field in the time domain is obtained by summing up all the frequency components, $\mathbf{E}_s(r,t) = Re\left(\int_{-\infty}^{\infty} d\mathbf{E}_s(r,\omega) \exp(-i\omega t)\right)$. The latter is conveniently expressed with a QNM expansion [3,13]

$$\mathbf{E}_s(r,t) = Re\left(\sum_m \beta_m(t)\widetilde{\mathbf{E}}_m(r)\right), \tag{3}$$

with

$$\beta_m(t) = \int_{-\infty}^{\infty} \frac{\omega \exp(-i\omega t)}{\widetilde{\omega}_m - \omega} \left(\int_{V_r} \Delta\varepsilon(r) \bar{\mathbf{E}}_b(r,\omega) \cdot \widetilde{\mathbf{E}}_m d^3 r\right) d\omega. \tag{4}$$

Replacing the spectral component $\bar{\mathbf{E}}_b(r,\omega)$ by the Fourier transform of the wavepacket $\mathbf{E}_b(r,t)$ (please note that Fourier transforms are represented with a horizontal bar), we obtain

$$\beta_m(t) = \int_{-\infty}^{\infty} \frac{\omega \exp(-i\omega t)}{2\pi(\widetilde{\omega}_m - \omega)} \left(\int_{V_r} \Delta\varepsilon \left(\int_{-\infty}^{\infty} \mathbf{E}_b(r,t') \exp(i\omega t')\, dt'\right) \cdot \widetilde{\mathbf{E}}_m d^3 r\right) d\omega. \tag{5}$$

Following [10], we introduce the overlap $O_m(t) = \int_{V_r} \Delta\varepsilon(r) \mathbf{E}_b(r,t) \cdot \widetilde{\mathbf{E}}_m d^3 r$. Equation (5) becomes

$$\beta_m(t) = \frac{1}{2\pi} \int_{-\infty}^{\infty} \frac{\omega}{\widetilde{\omega}_m - \omega} \left(\int_{-\infty}^{\infty} O_m(t') \exp(i\omega(t'-t))\, dt'\right) d\omega, \tag{6}$$

which coincides with Eq. (5) in [10].

In [10], it is shown that

$$\beta_m(t) = \frac{1}{2\pi} \int_{-\infty}^{\infty} \left(\int_{-\infty}^{\infty} \frac{\omega}{(\widetilde{\omega}_m - \omega)} O_m(t') \exp(i\omega(t'-t)) dt'\right) d\omega$$
$$= -O_m(t) + i\widetilde{\omega}_m \int_{-\infty}^{t} O_m(t') \exp(i\widetilde{\omega}_m(t'-t))\, dt'. \tag{7}$$

However, while we will see that this equality applies to a broad spectrum of complex functions $O_m(t')$, the proof is deficient in rigor, as we delineate in the subsequent Section.

## 3. Critical assessment of the proof in [10]

Complex analysis and generalized functions are used to derive Eq. (7) in [10]. While the equality holds, we have identified that several steps lack mathematical rigor in our scrutiny of the proof.

**Unjustified use of Fubini theorem.** The expression in Eq. (6) contains two integrals that have to be calculated sequentially, first considering the integral on $t'$ and then on $\omega$. In [10], the order of integration is inverted, implicitly using the Fubini theorem, whose assumptions are not satisfied. Indeed, one can first observe that an application of the Fubini-Tonelli theorem to the absolute value of the integrand shows that the latter is not integrable over $(-\infty, \infty)^2$ although $O_m$ is assumed to be in $L^1(\mathbb{R})$, the space of functions integrable over the real line. Second, it can be checked that the integral $\int_{-\infty}^{\infty} \frac{\omega}{\widetilde{\omega}_m - \omega} \exp(i\omega(t'-t)) d\omega$ diverges (see below), which implies that the Fubini theorem does not apply here.

**Divergent integrals.** We further revisit the computation of the integral with respect to $\omega$ in [10], where it is "shown" that for $t \neq t'$

$$\int_{-\infty}^{\infty} \frac{\omega}{\widetilde{\omega}_m - \omega} \exp(i\omega(t'-t)) d\omega = 2i\pi\widetilde{\omega}_m \exp(i\widetilde{\omega}_m(t'-t)) H(t'-t), \quad (8)$$

where H stands for the standard Heaviside function.

To derive Eq. (7), in [10], the residue theorem is applied twice on two semicircles in the upper and lower half of the complex plane (we refer here to the calculation of terms A and C in Eq. (6) in [10] with the red and green arcs shown in Fig. 1 in [10]). We believe that the derivation is incorrect.

Our primary concern lies in the fact that, for $\omega$ running over the green or red arcs, it is not consistently true that $Im(\omega) \to \infty$ (consider instances where $\omega$ is close to or at a specific distance from the blue diameter/line in Fig. 1). Thus, the contributions of the green and red integrals over the arcs cannot be assumed to be null.

Actually, $\int_{-\infty}^{\infty} \frac{\omega}{\widetilde{\omega}_m - \omega} \exp(i\omega(t'-t)) d\omega$ diverges for any real $t, t'$. To show that, we use the equality $\frac{\omega}{\widetilde{\omega}_m - \omega} = -1 + \frac{\widetilde{\omega}_m}{\widetilde{\omega}_m - \omega}$ and obtain

$$\int_{-\infty}^{\infty} \frac{\widetilde{\omega}_m}{\widetilde{\omega}_m - \omega} \exp(i\omega(t'-t)) d\omega = -\int_{-\infty}^{\infty} \exp(i\omega(t'-t)) d\omega + \int_{-\infty}^{\infty} \frac{\widetilde{\omega}_m}{\widetilde{\omega}_m - \omega} \exp(i\omega(t'-t)) d\omega. \quad (9)$$

The second integral on the right-hand side is defined and is equal to

$$\int_{-\infty}^{\infty} \frac{\widetilde{\omega}_m}{\widetilde{\omega}_m - \omega} \exp(i\omega(t'-t)) d\omega = 2i\pi\widetilde{\omega}_m \exp(i\widetilde{\omega}_m(t'-t)) H(t'-t). \quad (10)$$

The derivation of Eq. (10) can be accomplished by employing a methodology akin to that outlined in [10] utilizing the residue theorem. By employing the same integration contours depicted in Fig. 1 in [10], this time, the integrand tends to 0 as the radius of each semicircle goes to $\infty$, simply because $\lim_{|\omega| \to \infty} \left|\frac{\widetilde{\omega}_m}{\widetilde{\omega}_m - \omega}\right| = 0$ (which is not the case for $\left|\frac{\omega}{\widetilde{\omega}_m - \omega}\right|$).

The comparison between Eqs. (8) and (10) is eloquent. Furthermore, we observe that the first integral on the right-hand side of Eq. (9), $\int_{-\infty}^{\infty} \exp(i\omega(t'-t)) d\omega$, is divergent (or not defined) for any real $t, t'$. As the second integral in this equation is defined, see Eq. (10), the integral on the left-hand side, $\int_{-\infty}^{\infty} \frac{\widetilde{\omega}_m}{\widetilde{\omega}_m - \omega} \exp(i\omega(t'-t)) d\omega$, is also divergent.

## 4. Demonstration of Eq. (7)

In this Section, we provide a demonstration of Eq. (7) that does not rely on complex analysis. By introducing the spectral overlap $\bar{O}_m(\omega) = \int_{V_r} \Delta\varepsilon(r) \bar{\mathbf{E}}_b(r, \omega) \cdot \tilde{\mathbf{E}}_m d^3r$, Eq. (2) becomes

$$-i\omega\alpha_m(\omega) = -i\widetilde{\omega}_m \alpha_m(\omega) + i\omega\bar{O}_m(\omega). \quad (11)$$

We take the Fourier transform of Eq. (11) to move in the temporal domain

$$\int_{-\infty}^{\infty} -i\omega \alpha_m(\omega) \exp(-i\omega t)\, d\omega = -i\tilde{\omega}_m \int_{-\infty}^{\infty} \alpha_m(\omega) \exp(-i\omega t)\, d\omega$$
$$+ \int_{-\infty}^{\infty} i\omega \bar{O}_m(\omega) \exp(-i\omega t)\, d\omega. \tag{12}$$

To obtain Eq. (12), we need two assumptions on the driving pulse

- the Lebesgue integral of $\omega \alpha_m(\omega)$ is finite, i.e. $\int_{-\infty}^{\infty} |\omega \alpha_m(\omega)|\, d\omega < \infty$. (13a)

- the Lebesgue integral of $\omega O_m(\omega)$ is finite, i.e. $\int_{-\infty}^{\infty} |\omega \bar{O}_m(\omega)|\, d\omega < \infty$. (13b)

Let us consider the left-hand term: $\int_{-\infty}^{\infty} -i\omega \alpha_m(\omega) \exp(-i\omega t)\, d\omega$. It is equal to $\int_{-\infty}^{\infty} \alpha_m(\omega) \frac{d}{dt}\exp(-i\omega t)\, d\omega = \frac{d}{dt}\int_{-\infty}^{\infty} \alpha_m(\omega) \exp(-i\omega t)\, d\omega = \frac{d}{dt}\beta_m(t)$. We can do exactly the same for the last term of Eq. (12), $\int_{-\infty}^{\infty} i\omega \bar{O}_m(\omega) \exp(-i\omega t)\, d\omega = -\frac{d}{dt}O_m(t)$. Then, directly from Eq. (12), we have

$$\frac{d\beta_m(t)}{dt} = -i\tilde{\omega}_m \beta_m(t) - \frac{d}{dt}O_m(t), \tag{14}$$

which corresponds to the main result obtained for non-dispersive materials in [10], specifically the EME evolution. It is straightforward to verify that the expression for $\beta_m(t)$ provided in Eq. (7) satisfies Eq. (14). This completes the proof of Eq. (7).

In Appendix A, we present an alternative demonstration based on less stringent (possibly minimal) assumptions regarding the driving pulse: $\int_{-\infty}^{\infty} |\alpha_m(\omega)|\, d\omega < \infty$ and $\int_{-\infty}^{\infty} |\bar{O}_m(\omega)|\, d\omega < \infty$.

## 5. Clarification of the origin of the temporal derivative in the driving force

This section emphasizes the approximation made in classical CMT, where the driving force is assumed to be proportional to the incident field rather than its time derivative.

To clarify the approximation introduced in the CMT, we remember that a derivative in the real temporal domain amounts to multiply by the conjugated variable in the frequency domain in Fourier (or Laplace) analysis. We thus intuitively multiply the expression of the excitation coefficient by $\tilde{\omega}_m/\omega$ to remove the $\omega$-dependence in the numerator of Eq. (2). We obtain an approximate expression of the excitation coefficient denoted by $\alpha_m^{(app)}$

$$\alpha_m^{(app)}(\omega) = \frac{\tilde{\omega}_m}{\omega}\alpha_m = \frac{\tilde{\omega}_m}{\tilde{\omega}_m - \omega}\bar{O}_m(\omega). \tag{15}$$

We expect the approximate expression to be most accurate when the frequency of the monochromatic incident field is close to the QNM resonance frequency. As demonstrated in Section 4, the temporal excitation coefficient is given by $\beta_m^{(app)}(t) = i\tilde{\omega}_m \int_{-\infty}^{t} O_m(t') \exp\bigl(i\tilde{\omega}_m(t'-t)\bigr) dt'$. This leads by differentiation to

$$\frac{d\beta_m^{(app)}}{dt} = -i\tilde{\omega}_m \beta_m^{(app)}(t) + i\tilde{\omega}_m O_m(t), \tag{16}$$

in contrast to the EME equation

$$\frac{d\beta_m}{dt} = -i\tilde{\omega}_m \beta_m(t) - \frac{dO_m}{dt}. \tag{17}$$

Equation (16) aligns with the usual approach in CMT, here the driving force is proportional to an overlap integral between the incident field (rather than its derivative) and the QNM field, with a prefactor of $\tilde{\omega}_m$ as expected from dimensional analysis.

Next, we evaluate the accuracy of Eq. (16) for a 1D Fabry-Perot resonator in air. The resonator is illuminated at normal incidence by a 10-fs plane-wave Gaussian pulse (Fig. 1a). The central frequency of the pulse corresponds to the real part of the eigenfrequency of one of the QNMs (QNM 1 in Fig. 1b).

Figure 1c compares $|\beta_1|$ and $|\beta_2|$, computed using the approximate and EME equations. For each QNM, $O_m(t)$ is obtained by simply computing the overlap between the QNM electric field $\tilde{\mathbf{E}}_m(\mathbf{r})$ and the Gaussian pulse $\mathbf{E}_b(\mathbf{r},t)$: $O_m(t) = \int_{V_r} \Delta\varepsilon(\mathbf{r})\mathbf{E}_b(\mathbf{r},t)\cdot\tilde{\mathbf{E}}_m(\mathbf{r})d^3\mathbf{r}$. The values of $|\beta_1|$ and $|\beta_2|$ are then determined by solving the differential equations (16) and (17). Significant differences are observed primarily for the off-resonant QNM 2, for which the resonance frequency differs from the pulse central frequency – the approximation made in Eq. (15) by multiplying by the excitation coefficient by $\tilde{\omega}_m/\omega$ becomes inaccurate. The reverse occurs when we adjust the central frequency of the incident pulse to match $Re(\tilde{\omega}_2)$, implying the QNM field distribution does not play any role in the difference reported. Additionally, note that in classical CMT, the coupling coefficient is typically fitted, and more precise predictions than those derived from Eq. (17) are possible, as discussed in Section 3 of [10].

Eq. (16): $\dfrac{d\beta_m^{(app)}}{dt} = -i\tilde{\omega}_m \beta_m^{(app)}(t) + i\tilde{\omega}_m O_m(t)$

Eq. (17): $\dfrac{d\beta_m}{dt} = -i\tilde{\omega}_m \beta_m(t) - \dfrac{dO_m(t)}{dt}$

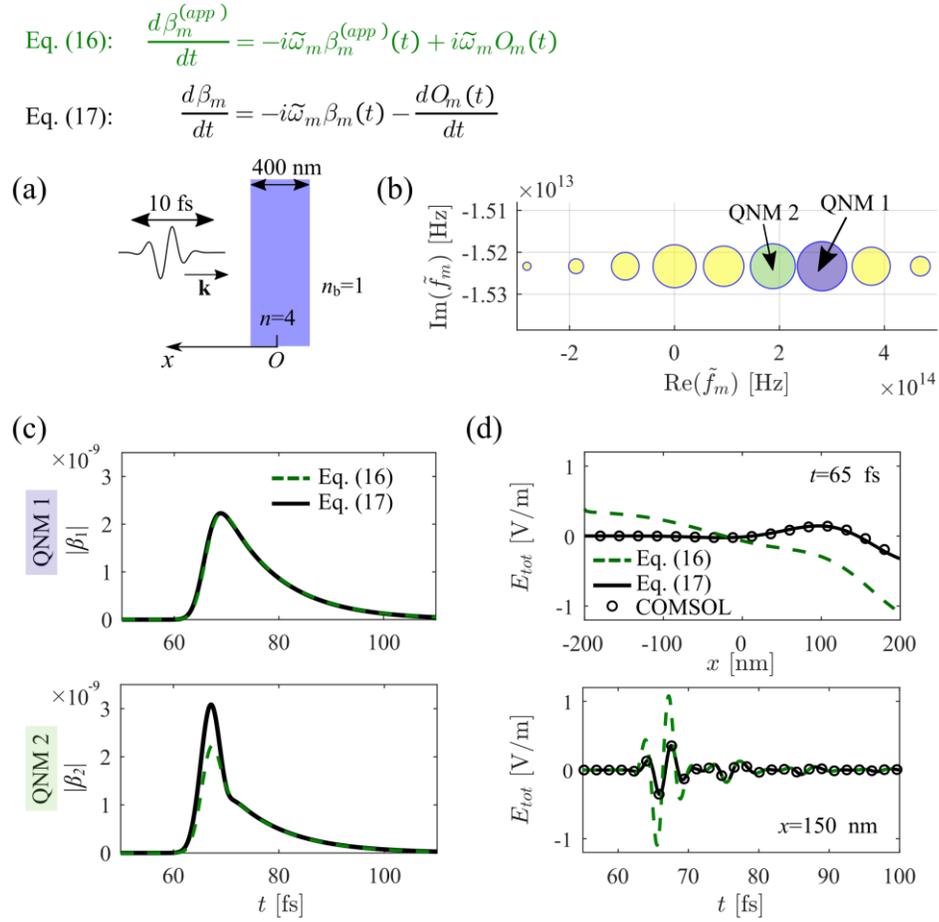

**Fig. 1.** Comparison of the EME and approximate equations for a 1D Fabry-Perot resonator illuminated by a short 10-fs pulse. **(a)** Sketch of the geometry. **(b)** QNMs of the resonator in the complex frequency plane, calculated using the MAN software. The circle size indicates the excitation level of the QNMs. The real part of the eigenfrequency for QNM 1 corresponds to the central frequency of the incident pulse. **(c)** The values of $\beta_1$ and $\beta_2$ obtained from both the EME and approximate equations. **(d)** A comparison of the total-field reconstruction using the EME and approximate equations with 250 QNMs against full-wave numerical results from COMSOL.

Temporal excitation coefficients, being abstract quantities, cannot be directly measured. Figure 1d offers a comparison using measurable quantities, such as the total field. The two panels in Fig. 1d show the reconstructed total field: first, inside the resonator for a fixed time $t = 65$ fs, and second, at a fixed coordinate $x = 150$ nm as a function of time. For the reconstruction, 250 QNMs are used in the expansion of Eq. (1). The EME equation (17) yields a reconstruction that perfectly matches reference data from COMSOL time-domain simulations, while the approximate equation (16) exhibits significant deviations.

It is important to note that the driving pulse is ultra-short and has a broad spectral range. Furthermore, all the QNMs have low quality factors, making them easily excited even when the central frequency of the driving pulse deviates significantly from the QNM resonance frequency. This is why we used 250 QNMs in the reconstruction shown in Fig. 1d. With such a large number of QNMs, the maximum difference between the black curves of Eq. (17) and the COMSOL data is less than 0.005 for both curves in Fig. 1d, providing a strong quantitative validation of the approach.

## 6. Conclusion

The derivations of the EME equation in Section 4 and Appendix A are straightforward and do not depend on complex analysis or generalized functions, providing a solid foundation for the equation across a broad range of incident pulses.

The numerical tests in Section 5 are intentionally carried out on a simple 1D geometry, commonly used with CMT, which emphasizes the validity of the EME equation. This is achieved by using a driving term $-\int_{V_r} \Delta\varepsilon(\mathbf{r}) \frac{d\mathbf{E}_b(\mathbf{r},t)}{dt} \cdot \tilde{\mathbf{E}}_m d^3\mathbf{r}$ instead of $i\tilde{\omega}_m \int_{V_r} \Delta\varepsilon(\mathbf{r}) \mathbf{E}_b(\mathbf{r},t) \cdot \tilde{\mathbf{E}}_m d^3\mathbf{r}$ when $\Delta\varepsilon(\mathbf{r})$ is non-dispersive. As noted in [10], under the slowly-varying envelope approximation, $\frac{d\mathbf{E}_b(\mathbf{r},t)}{dt}$ can often be replaced by $-i\omega \mathbf{E}_b(\mathbf{r},t)$, where $\omega$ is the carrier frequency of the driving pulse. This suggests that fitting a coupling constant with a driving term proportional to $\mathbf{E}_b(\mathbf{r},t)$ will yield accurate predictions when $\tilde{\omega}_m \approx \omega$, as demonstrated in Section 5.

Nonetheless, many modern electromagnetic software tools, including general solvers like COMSOL and specialized options [14-16] now compute QNMs. Additionally, QNM normalization has been mastered and is implemented in specialized software [16]. As a result, the time-derivative-based analytical expression, $-\int_{V_r} \Delta\varepsilon(\mathbf{r}) \frac{d\mathbf{E}_b(\mathbf{r},t)}{dt} \cdot \tilde{\mathbf{E}}_m d^3\mathbf{r}$, can be efficiently calculated as a spatial overlap integral. Therefore, we anticipate that CMT models without fitting parameters will be readily used to interpret both experimental and numerical results. Notably, the temporal toolbox in [16] effectively addresses the exact Maxwell evolution equation, and efforts are already underway to model the second CMT equation for port coupling [13,16-18].

## Appendix A: Demonstration of Eq. (7) via Fourier inversion

In this Appendix, we again aim at calculating (see Eq. 7)

$$\beta_m(t) = \frac{1}{2\pi} \int_{-\infty}^{\infty} \left( \int_{-\infty}^{\infty} \frac{\omega}{(\tilde{\omega}_m - \omega)} O_m(t') \exp(i\omega(t' - t)) dt' \right) d\omega, \qquad (A1)$$

with minimum assumption on the incident field, i.e. on $O_m(t')$. The approach just assumes that $O_m$ is in $L^1(\mathbb{R})$, the space of functions integrable over the real line. It is based on the well-known Fourier inversion formula.

For fixed values of $\omega$, we first calculate the integral $\frac{1}{2\pi} \int_{-\infty}^{\infty} \frac{\omega}{(\tilde{\omega}_m - \omega)} O_m(t') \exp(i\omega(t' - t)) dt'$, which is equal to $\frac{\omega}{(\tilde{\omega}_m - \omega)} \exp(-i\omega t) \bar{O}_m(\omega)$. Further integrating over $\mathbb{R}$ with respect to $\omega$, we aim at calculating $\int_{-\infty}^{\infty} \frac{\omega}{(\tilde{\omega}_m - \omega)} \exp(-i\omega t) \bar{O}_m(\omega) d\omega$. Replacing $\frac{\omega}{(\tilde{\omega}_m - \omega)}$ by $\frac{\tilde{\omega}_m}{(\tilde{\omega}_m - \omega)} - 1$, we first obtain $\beta_m(t) = \int_{-\infty}^{\infty} \left( -1 + \frac{\tilde{\omega}_m}{(\tilde{\omega}_m - \omega)} \right) \exp(-i\omega t) \bar{O}_m(\omega) d\omega$ and thus find

$$\beta_m(t) = -O_m(t) + \widetilde{\omega}_m \int_{-\infty}^{\infty} \frac{1}{(\widetilde{\omega}_m - \omega)} \exp(-i\omega t)\, \bar{O}_m(\omega) d\omega. \tag{A2}$$

To evaluate the integral on the right-hand side, we demonstrate that the function $G(t) = \exp(i\widetilde{\omega}_m t) F(t)$, where $F(t) = \int_{-\infty}^{\infty} \frac{1}{(\widetilde{\omega}_m - \omega)} \exp(-i\omega t)\, \bar{O}_m(\omega) d\omega$ satisfies a simple differential equation that we can solve. It can be verified that $G$ is a differentiable function of the variable $t \in \mathbb{R}$, because the conditions of the well-known differentiation theorem for integrals are met for functions in $L^1(\mathbb{R})$. Differentiating $G$ with respect to $t$ yields $\frac{dG}{dt} = i \exp(i\widetilde{\omega}_m t) \int_{-\infty}^{\infty} \exp(-i\omega t)\, \bar{O}_m(\omega) d\omega = i \exp(i\widetilde{\omega}_m t) O_m(t)$. Considering that $\lim_{t \to \infty} G(t) = 0$ (the limit and the integral can be interchanged due to the integrand being suitably dominated), we obtain $G(t) = i \int_{-\infty}^{t} \exp(i\widetilde{\omega}_m t')\, O_m(t') dt'$.

Thus $F(t) = i \int_{-\infty}^{t} \exp(i\widetilde{\omega}_m(t' - t))\, O_m(t') dt'$ and coming back to Eq. (A2), we find

$$\beta_m(t) = -O_m(t) + i\widetilde{\omega}_m \int_{-\infty}^{t} \exp(i\widetilde{\omega}_m(t' - t))\, O_m(t') dt'. \tag{A3}$$

This concludes a demonstration of Eq. (7).